\documentclass{aa}  
\usepackage{graphicx}
\usepackage{natbib}

\usepackage{txfonts}
\begin{document}
 \title{X-ray emission from the M9 dwarf 1RXS J115928.5-524717}
   \subtitle{Quasi-quiescent coronal activity at the end of the main sequence}

   \author{J. Robrade
          \and
          J.H.M.M. Schmitt
          }


        \institute{Universit\"at Hamburg, Hamburger Sternwarte, Gojenbergsweg 112, D-21029 Hamburg, Germany\\
       \email{jrobrade@hs.uni-hamburg.de}
             }

   \date{Received 27 October 2008; Accepted 19 December 2008}

 
  \abstract
{}
{X-ray emission is an important diagnostic to study magnetic activity in presumably fully convective, very low-mass stars with
an effectively neutral photosphere.}
{We analyse an XMM-Newton observation of 1RXS J115928.5-524717, an ultracool dwarf with spectral type M9 and compare
its X-ray properties to those of other similar very late-type stars.}
{We clearly detected 1RXS J115928.5-524717 at soft X-ray energies in all EPIC detectors. 
Only minor variability was present during the observation and we attribute the X-ray emission to quasi-quiescent activity. 
The coronal plasma is described well by a two-temperature model at solar metallicity with
temperatures of 2\,MK and 6\,MK and an X-ray luminosity of about $L_{\rm X} = 1.0 \times 10^{26}$~erg/s in the 0.2\,--2.0\,keV band.
The corresponding activity level of log~$L_{\rm X}$/$L_{\rm bol}\approx -4.1$ points to a moderately active star. Altogether, X-ray activity from very low-mass stars
shows similar trends to more massive stars, despite their different interior structure.}
{1RXS~J115928.5-524717 is, after LHS 2065, the second ultracool M9 dwarf that emits X-rays at detectable levels in quasi-quiescence. While faint in absolute numbers,
both stars are rather X-ray active, implying the existence of an efficient dynamo mechanisms that is capable of creating magnetic activity and coronal X-ray emission. }
   \keywords{Stars: activity -- Stars: coronae -- Stars: individual 1RXS J115928.5-524717 -- Stars: low-mass, brown dwarfs -- X-rays: stars
               }

   \maketitle
%

\section{Introduction}

The ultracool dwarf \object{1RXS J115928.5-524717} with a spectral type of M9V is a sparsely studied, very low-mass star in the solar vicinity.
The variable RASS (ROSAT All Sky Survey) source 1RXS~J115928.5-524717 was first investigated by \cite{grei00} in a search
for GRB X-ray afterglow candidates.
Later, \cite{ham04} associated 1RXS~J115928.5-524717 with the nearby late-type star 2MASS J11592743-5247188.
During a strong flare observed in 1991 with ROSAT, its X-ray activity level reached a value of log~$L_{\rm X}$/$L_{\rm bol} = -1$, one of the highest levels observed so far in any star.
\cite{ham04} initiated optical follow-up observations, investigated archival data in the optical/IR band, and concluded that 1RXS~J115928.5-524717 is a high proper-motion star
with spectral type M9\,$\pm 0.5$ at a distance of only $11 \pm 2$~pc and a total luminosity of $L_{\rm bol}= 1.2\times 10^{30}$~erg/s. 
Observations with VLT/UVES
did not show the presence of Li and indicated a relatively fast rotation of $V sini \approx$~25~km/s \citep{ham05}.
Recently, \cite{pha08} described the same star as DENIS-P J115927.4-524718 with a spectral type again of M9 and derived a distance of 10.2\,$\pm$\,1.7~pc. 
Obviously in the solar vicinity the populations of older, very low-mass stars and young brown dwarfs overlap in the regime of late M~dwarfs 
\citep[see e.g.][]{giz00}. Therefore a definite assignment to a specific population for an individual object like 1RXS~J115928.5-524717, based on IR-magnitudes alone, is difficult. 

The presence of magnetic activity phenomena in the outer atmospheric layers of these ultracool stars is remarkable, and
among other diagnostics, X-ray emission can put important constraints on the possible activity generating mechanisms.
Very low-mass stars are generally assumed to be fully convective and hence a solar-type dynamo is not expected to operate.
In more massive, solar-type stars magnetic activity is generated
at the interface layer between the radiative core and outer convection zone ($\alpha \Omega$ dynamo). However, around spectral type mid-M, the stellar interior  
becomes fully convective and the presence of magnetic activity in these objects requires another dynamo mechanisms, e.g. an $\alpha^{2}$ or turbulent dynamo, to operate. 
Furthermore, the cool photospheres with temperatures of $T_{eff} \lesssim 2500$\,K should be effectively neutral, 
leading to a high electric resistivity, inhibiting the transport of magnetic energy through the photosphere
and consequently inhibiting magnetic activity in the outer layers of the star.
A comprehensive overview of the theory of very low-mass stars is given e.g. in \cite{cha00a}.

Evidence for magnetic activity in the X-ray regime comes from large flares detected from several very low-mass stars, 
but also from quiescent emission that points to the existence of stable coronae as observed in solar-like stars.
Other indicators of magnetic activity are most prominently their often strong and variable H${\alpha}$ emission \citep{giz00, moh03, schmidt07},
suggesting the ubiquitous existence of a chromosphere and of radio gyrosynchrotron emission, which requires the presence of fast electrons and magnetic fields \citep{ber02}.
Recently, strong magnetic fields with $Bf$-values of up to a few kG have been detected on several ultracool dwarfs,
utilizing the profiles of their FeH lines \citep{rei07}.

The X-ray source 1RXS~J115928.5-524717 has not been studied in detail up to now. Only the ROSAT X-ray detection is mentioned in the 
literature, and the X-ray signal was rather weak during the observation, despite the observed large flare.
In total 52 (source+background) photons were detected in 356\,s observation time divided into 17 exposures, 
whereas roughly 24 background photons were expected.
Nearly all source photons were detected in two subsequent exposures separated by 1.6\,h in a total observation time of less than a minute. 
The X-ray flare decayed rather fast with a decay  time (e-folding) of about 1.5\,h.
From the spectral modelling of this quite limited signal they deduced a plasma temperature of about 2.5\,MK, albeit with a large error.
A peak X-ray luminosity of $L_{\rm X}\approx1.3 \times 10^{29}$~erg/s and 
an upper limit of the quiescent flux of $L_{\rm X}\lesssim 1.8\times 10^{27}$~erg/s were derived.
Little is known about the magnetic activity on 1RXS~J115928.5-524717; however, together with the M9 dwarf LHS~2065, 
whose quiescent X-ray emission was recently confirmed by {\it XMM-Newton} \citep{rob08a}, it is among the latest stars detected in X-rays.

The X-ray observations of ultracool dwarfs are quite rare, and we examined a previously unpublished
{\it XMM-Newton} observation of the M9 dwarf 1RXS~J115928.5-524717.
Here we report its X-ray detection in the quasi-quiescent state with the EPIC detectors
at soft X-ray energies despite rather unfavourable background conditions.
In Sect.\,\ref{ana} we describe the observation and data analysis, in Sect.\,\ref{res} we present our results, put them in the context of similar objects in Sect.\,\ref{act}, 
and summarise our findings in Sect.\ref{sum}.

\section{Observations and data analysis}
\label{ana}

1RXS~J115928.5-524717 was observed by {\it XMM-Newton} in January 2006 for approximately 38\,ks (Obs.-ID 0301430101).
We only consider data taken with the EPIC (European Photon Imaging Camera) detector, i.e., the PN and MOS, which were both
operated in 'Full Frame' mode with the thin filter inserted. No useful source signal is present in the RGS data and
the OM operated in the imaging mode with the UVW1 filter ($\lambda_{eff}$\,=\,2910\,\AA), but we did not detect 1RXS~J115928.5-524717.
The XMM-Newton data analysis was carried out with the Science Analysis System (SAS) version~8.0 \citep{sas}.
The background conditions of this data are unfavourably high; however, only a few minor data gaps (especially around the observation time of 12\,ks) are present.

\begin{figure}[ht]
\hspace*{0.3cm}
\includegraphics[width=82mm]{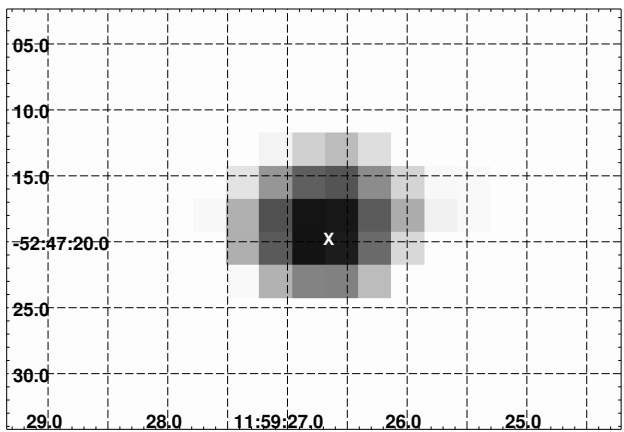}
\caption{\label{im}Image of the detected X-ray source and expected position of 1RXS~J115928.5-524717 as marked by X (see text).}
\end{figure}

We applied standard selection criteria, but due to the overall high background levels, we used
no time filtering for source detection purposes. Instead, we suppressed the background by restricting the energy range of the considered photons 
since we expect a possible source signal to be predominantly at soft energies; specifically we used the 0.2\,--\,1.0~keV band where an adequate S/N is present in all detectors.
From these photons we created images for each detector
and then ran the source detection algorithm 'edetect\_chain', requiring a maximum likelihood of at least 10. The source is clearly detected in all EPIC detectors.
Figure\,\ref{im} shows a partial image obtained in the MOS1 detector that provides the highest spatial resolution,
covering the detected X-ray source and the expected position of 1RXS~J115928.5-524717.

To optimize the SNR we constrained our further analysis to the core of the PSF. For the generation of light curves we merged all EPIC data 
and extracted the source photons from a 10\,\arcsec 
circular region around the position of 1RXS~J115928.5-524717. The background was taken from a nearby region. 
For a spectral analysis the spectra from the individual instruments were considered, but since
the signal is rather weak in the MOS detectors, we only used the data from the PN detector. 
The PN detector has by far the highest sensitivity and the energy range with sufficient S/N is widest.
To obtain a source spectrum, we again used only photons from a 10\,\arcsec circular region around the source position 
and determined the background from a nearby region on the same CCD.
Spectral analysis was performed with XSPEC~V12.3 \citep{xspec}, and we
used multi-temperature APEC models with solar abundances as given by \cite{grsa}. 
We note that the applied metallicity is interdependent with the emission measure and different
combinations of both parameters lead to very similar results. Due to the proximity of the target, interstellar absorption is negligible and
not required in the modelling of the X-ray data.

To study a possible remaining background contamination of the analysed data,
we also extracted photons from a reduced dataset that excludes periods of very high background (PN rate above 10\,keV $>$~3~cts/s; duration 28.8\,ks) 
or used different source and background extraction regions and crosschecked our findings. 
We find overall consistent results that only marginally depend on the specific combination of datasets.

\section{Results}
\label{res}

\subsection{Source detection and light curves}

A faint X-ray source is clearly detected close to the expected position of 1RXS~J115928.5-524717 in all EPIC detectors.
The detection at soft X-ray energies is very robust, in the most sensitive instrument PN alone the detected excess has a significance of more than 8~$\sigma$.
We note that the source is not significantly detected at higher energies, e.g. in the 1.0\,--\,3.0~keV band, even with a factor of three reduced minimum likelihood.
The source detection parameter for the individual instruments are summarised in Table\,\ref{log}, denoted errors are Poissonian errors (1\,$\sigma$).

\begin{table} [ht!]
\setlength\tabcolsep{5pt}
\begin{center}
\caption{\label{log} X-ray detection of 1RXS~J115928.5-524717 with the SAS-tool 'edetect\_chain', optical pos: RA: 11 59 26.64  DEC: -52 47 19.7.}
\begin{tabular}{l|rrr}
\hline
Par.  & PN & MOS1 & MOS2 \\\hline
On Time (ks) & 36.0 & 38.1 & 38.1  \\
RA~~~~~~~11 59 &  26.84($\pm$\,1.0\,\arcsec)  & 26.83  ($\pm$\,1.7\,\arcsec)& 26.73  ($\pm$\,1.7\,\arcsec) \\
DEC~~ -52 47 &  20.25($\pm$\,1.0 \arcsec) & 18.79 ($\pm$\,1.7\,\arcsec)& 19.01 ($\pm$\,1.7\,\arcsec) \\
Counts & 177$\pm$ 20 & 51 $\pm$11& 43 $\pm$9 \\
Det. likelihood & 78 & 29 & 23 \\\hline\hline
\end{tabular}
\end{center}
\end{table}

The derived source positions from the EPIC detectors agree with each other within errors and are 
within 2\,\arcsec~of the expected position of 1RXS~J115928.5-524717 for epoch 2006.0; 
we derived the expected position from the 2MASS coordinates (11 59 27.43 -52 47 18.8 at epoch 1999.36) 
 - which are more accurate than the ROSAT coordinates  -  and the proper motion 
(-1077\,mas/yr, -131\,mas/yr) as given by \cite{ham04}, leading to an expected position of RA: 11 59 26.64 and DEC: -52 47 19.7.
The detected source is the only X-ray source within 5\,\arcmin~from the on-axis position.
No other known X-ray source of comparable strength is located in the vicinity of 1RXS~J115928.5-524717, 
therefore the identification is unambiguous, and the observed soft X-ray spectra make an unknown extragalactic source extremely unlikely.

\begin{figure}[ht]
\includegraphics[width=93mm]{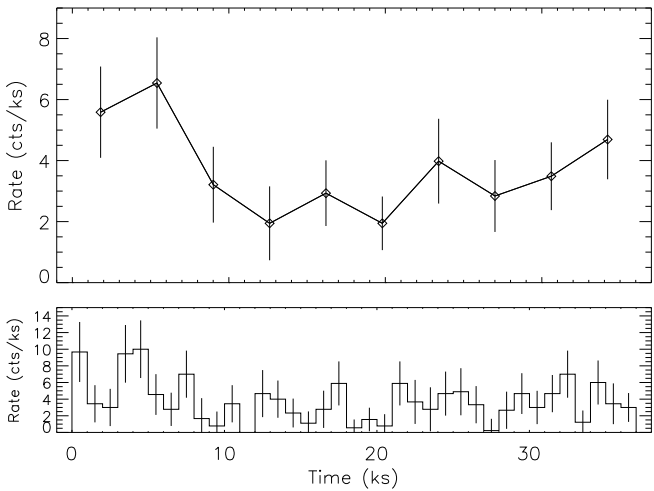}
\caption{\label{lc}Light curve of 1RXS~J115928.5-524717 derived from the combined EPIC data with a binning of 1\,h (top) and 1\,ks (bottom).}
\end{figure}

To investigate X-ray variability of 1RXS~J115928.5-524717
we created light curves with 1\,ks and 1\,h binning respectively from the photons detected in a 10\,\arcsec region around the source position.
In Fig.\,\ref{lc} we show the thus obtained, background subtracted light curve derived from the merged EPIC data in the 0.2\,--\,1.0~keV band for two different time resolutions.

The light curves shown in Fig.\,\ref{lc} clearly confirm that the detection of 1RXS~J115928.5-524717 with {\it XMM-Newton} is not caused
by a single flare event,
rather persistent X-ray emission is detected during the total observation, however, variability is also present at a significant level.
Therefore, instead of the commonly used term quiescence, we attribute the X-ray emission to a quasi-quiescence flux level. 
Some enhanced activity or a smaller flare might be present at the beginning of the observation, but the overall
variations in X-ray brightness appear quite smooth, at least when the 1\,h averaged light curve shown in the upper panel is considered. 
Otherwise, the light curve in lower panel with 1\,ks time bins indicates stronger variability, i.e. frequent smaller flares of different amplitude and duration.
Given the errors, both scenarios are possible and
the maximum X-ray variability may be around or even less than a factor of two, but could also be easily of the order of a few as suggested from light curves with shorter time bins.

We searched for spectral variations related to changes in X-ray brightness for the 1\,h time bins by studying the respective hardness ratio
HR\,=\,(H\,-\,S)/(H\,+\,S) with S\,=\,0.2\,--\,0.6~keV and H\,=\,0.6\,--\,1.0~keV being the photon energy bands.
The SNR is rather poor and we find no clear correlation between HR and count rate,
linear regression resulted in a slope of $0.035 \pm 0.044$. Therefore we performed the spectral analysis for the total observation.

\subsection{Spectral analysis}

\begin{figure}[ht]
\includegraphics[width=54mm,angle=-90]{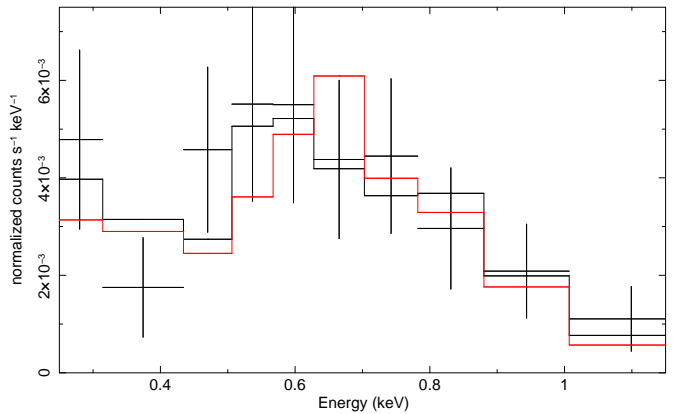}
\caption{\label{spec}PN spectrum of 1RXS~J115928.5-524717 with a binning of at least 15~cts/bin and the applied spectral models (black: 2-T, red: 1-T).}
\end{figure}

To determine the spectral properties of 1RXS~J115928.5-524717, we fitted the PN spectrum
with spectral models consisting of one and two temperature components. 
The elemental abundances cannot be constrained with the existing data and were set to solar values.
We show the X-ray spectrum and both respective best fit models in Fig.\,\ref{spec}, the derived spectral properties are summarised in Table\,\ref{fit}. 
The model with one temperature component is technically acceptable given the errors, but results in an somewhat poorer fit.
As visible in Fig.\,\ref{spec}, greater discrepancies are present for this model especially around the peak of the spectrum, 
indicating that it is a oversimplification. Given the fact that a two-temperature component model is also physically more realistic, 
since X-ray spectra with higher SNR generally require multiple components, we adopt its results for further discussion.

The best fit two-temperature model corresponds to a source flux of  $6.9 \times 10^{-15}$~erg\,cm$^{-2}$\,s$^{-1}$
in the 0.2\,--\,2.0\,~keV band; for the ROSAT 0.1\,--\,2.4\,keV band we derive a roughly 15\% higher flux of $8.0 \times 10^{-15}$~erg\,cm$^{-2}$\,s$^{-1}$.
Adopting a distance of 11\,pc for 1RXS~J115928.5-524717, we obtain an
X-ray luminosity of  $L_{\rm X}= 1.0 \times 10^{26}$~erg/s in the 0.2\,--\,2.0~keV band, 
corresponding to a moderate activity level of around log~$L_{\rm X}$/$L_{\rm bol} = -4.1$.
Our spectral modelling results in a cool plasma component with a temperature of 2\,MK and a moderately hotter component with temperatures around 6~MK.
The cooler plasma apparently slightly dominates the coronal emission measure distribution during these observation.
A significant, hot ($\gtrsim$10MK) component is not present as expected for quasi-quiescent emission of a moderately active star.
On the other hand, an inactive corona composed only of very cool (2\,MK) plasma as seen on the Sun during activity minimum can also be ruled out. 
We note that the coronal temperatures of 1RXS~J115928.5-524717 are comparable to the typical coronal temperatures derived for higher mass with a similar activity level. 

\begin{table} [ht!]
\begin{center}
\caption{\label{fit} Spectral results derived from the PN data.}
\begin{tabular}{lrrl}
\hline
Par.  & 2-T & 1-T & unit\\\hline
T1& 0.18$\pm$0.05 & 0.30$\pm$0.03 & keV\\
EM1&3.1$\pm1.3 \times 10^{48}$& 5.1$\pm0.7 \times10^{48}$&cm$^{-3}$ \\
T2&0.56$\pm$0.19 & - & keV\\
EM2& 2.3$\pm1.4 \times 10^{48}$& - & cm$^{-3}$\\\hline
$\chi^2_{red}${\tiny(d.o.f.)}& 0.70 (6) & 0.83 (8 )&\\
L$_{\rm X}$ (0.2\,--\,2.0~keV) &1.0\,$\times 10^{26}$ & 0.9\,$\times 10^{26}$ &erg/s  \\\hline\hline
\end{tabular}
\end{center}
\end{table}

The spectra of magnetically active stars are usually dominated by emission lines, which remain unresolved in the PN spectra.
For moderately active stars like 1RXS~J115928.5-524717, we expect a cool plasma component emitting strong lines of \ion{O}{vii} 
with a peak formation temperature of 2\,MK and of \ion{O}{viii}, which is predominantly formed at 3\,--\,5~MK.
The He-like triplet lines of \ion{O}{vii} at energies around 570\,eV and the \ion{O}{viii} Ly$\alpha$ line at 650\,eV then
naturally explain the peak of the spectrum around 0.6\,keV,
whereas the emission above 0.7\,keV requires the presence of hotter plasma that emits e.g. several strong \ion{Fe}{xvii} lines in this energy range.
Thus, the theoretical expectations combined with the observed spectral shape also favour the two-temperature model and indicate that residuals in the one-temperature model
are due to an underprediction of \ion{O}{vii} and an overprediction of \ion{O}{viii}, instead of noise or background contamination.

As an additional crosscheck we converted the observed count rates from each MOS and PN dataset
to an energy flux by using ECF's (Energy Conversion Factors) in the 'edetect\_chain' task. For the 0.2\,--\,1.0~keV band used in source detection, we respectively
obtain a source flux of $ 6.3 \pm 0.8 \times 10^{-15}$~erg\,cm$^{-2}$\,s$^{-1}$ (PN) and of $4.6~(5.1) \pm 1,3 \times 10^{-15}$~erg\,cm$^{-2}$\,s$^{-1}$ (MOS 1/2).
The MOS predicts a slightly lower flux; however given the errors, all methods agree rather well.
The values are also consistent with the corresponding source flux derived from the spectral modelling, 
e.g. $6.1 \times 10^{-15}$~erg\,cm$^{-2}$\,s$^{-1}$ for the two-temperature model in the 0.2\,--\,1.0~keV band.
We note that the given errors are pure statistical errors and further errors, e.g. in the distance of the star, are present and may affect derived values by almost up to 50\%.
The quiescent flux from 1RXS~J115928.5-524717 is an order of magnitude below and therefore fully consistent with the upper limit given by \cite{ham04} for the ROSAT data.

\section{X-ray activity in ultracool dwarfs}
\label{act}

In the following we put our results from 1RXS~J115928.5-524717 into the context of (quasi-) quiescent X-ray properties of other ultracool M dwarfs in the solar vicinity.
As already noted, the detection statistics are rather sparse for stars in the regime of spectral type beyond M7 and, additionally, 
we do not consider stars that are only detected during flares.
As mentioned above, age and mass are not known for all objects and therefore a clear distinction between stars and hot brown dwarfs is not possible \citep[see e.g.][]{giz00};
however, most of the objects discussed here are likely to be old ($\gtrsim$1.0~Gyr) and therefore real stars, as deduced 
e.g. from the absence of lithium absorption.

The latest stars detected by ROSAT in quasi-quiescence are the M7 dwarf VB~8  \citep{fle93} and - at least at selected time intervals - the M9 dwarf LHS~2065 \citep{schmitt02}.
More recently, two M9 dwarfs (LHS~2065, 1RXS~J115928.5-524717) and three M8 dwarfs (VB~10, LP~412-31, TVLM 513-46546)
have been clearly detected in X-rays in quasi-quiescence by {\it Chandra} and {\it XMM-Newton} as summarised in Table\,\ref{sam}. 
Besides emitting quiescent X-ray emission, all of them are also known to show
X-ray flares (and a possible flare from TVLM 513-46546) with flux increases up to a factor 100. 
We note that \cite{aud07} report an X-ray detection of the early L-dwarf binary Kelu-1 with {\it Chandra}, 
but a more quantitative analysis of these data was not possible since only four photons were detected in roughly six hours of
observation time.

\begin{table*}[t]
\setlength\tabcolsep{5pt}
\begin{center}
\caption{\label{sam}Ultracool M dwarfs with detected quasi-quiescent X-ray emission.}
\begin{tabular}{llrrrrrl}
\hline
Star  & Sp. type &V sin\,i & Bf&  log $L_{\rm H\alpha}$/$L_{\rm bol}$& log L$_{\rm X}$ & log $L_{\rm X}$/$L_{\rm bol}$& Ref.(X-ray) \\\hline
VB 10\,$^{a}$ & M8 &6& 1.3&  -4.4/-5.0 &25.4 & -4.9 & \cite{fle03}\,$^{\rm{1}}$\\
LP~412-31\,$^{a}$& M8 &9& $>$3.9&-3.9/-4.4& 27.2 & -3.1 & \cite{ste06}\\
TVLM 513-46546\,$^{b}$ & M8.5 &60 &-& -4.8/-5.0& 24.9 & -5.1 & \cite{ber08a}\\
LHS~2065\,$^{a}$ & M9 &12& $>$3.9&-4.0/-4.3&26.3 & -3.7 & \cite{rob08a}\,$^{\rm{2}}$\\
1RXS~J115928.5-524717\,$^{c}$ & M9 & 25 &-&-&26.0 & -4.1 & this work\\\hline\hline
\end{tabular}
\end{center}
\begin{list}{}{}
\item $^{a}$ additional values from \cite{rei07}, $^{b}$ from \cite{moh03}, $^{c}$ from \cite{ham05}
\item $^{\rm{1}}$ Re-observed by \cite{ber08b}, $^{\rm{2}}$ Detected at specific times also by \cite{schmitt02}
\end{list}
\end{table*}

Given the fact that fewer than ten photons were detected in the total observation of TVLM 513-46546 and that the 
quality of the LHS~2065 data is rather poor, only three stars remain for a spectral comparison. 
The quasi-quiescent state of VB~10 was modelled by \cite{fle03} with a 2.8\,MK plasma component (based on 26 photons).
The observation performed in 2007 was better described by a significantly hotter model; however, most of the photons in this observation were from a flare. 
The quasi-quiescent emission from LP~412-31 is not discussed in detail in \cite{ste06}.
To obtain a more complete picture from all available data, we derived its coronal properties in analogy to 1RXS~J115928.5-524717,
using a two-temperature model with solar metallicity. We find best-fit values around T1\,=\,3.5\,MK and T2\,=\,15\,MK with EM1/2$=4.8/3.3 \times10^{48}$~cm$^{-3}$,
though with substantial error, especially on the hotter component.
These values lead to X-ray luminosities of $L_{\rm X}= 1.3 \times 10^{27}$~erg/s in the 0.2\,--\,2.0~keV band, confirming that
LP~412-31 is an order of magnitude X-ray brighter and, correspondingly, roughly a factor ten more active. 
Its coronal temperatures are also higher than those of 1RXS~J115928.5-524717, and larger amounts of flaring ($\gtrsim 10$\,MK) plasma are apparently also present in quasi-quiescence.
In contrast, the corona of the less active star VB~10 seems to be cooler than those of the other stars. 
This indicates that coronal temperatures and X-ray activity also correlate in fully convective objects.
A counter-example is the correlation between radio and X-ray luminosity, known as the G\"udel-Benz relation \citep{gue93},
which is valid for a variety of magnetic activity phenomena in the F-type to mid-M type stars, but is strongly violated in the regime of ultracool dwarfs \citep{ber08b}.
While spectral information on ultracool dwarfs is still sparse in the X-ray regime, at least the studied objects show the same trend as observed for solar-like stars, 
i.e. hotter coronae in more active stars.

The X-ray activity levels for higher mass stars with spectral types F to mid-M span the range of log~$L_{\rm X}$/$L_{\rm bol} \approx -3 ...-7$, 
with the saturation limit around log~$L_{\rm X}$/$L_{\rm bol} = -3$ \citep[see e.g.][]{piz03}. This trend is most likely to continue all the way down the main-sequence as indicated by
the activity levels seen in very low-mass stars. The X-ray detected very low-mass stars already span two orders of magnitude in activity level,
whereas the highest activity levels in these objects are similar to those of higher mass stars. The absence of ultracool dwarfs with very low activity levels
is not surprising, since it would require a much higher sensitivity than available in the performed X-ray observation. 

In H$\alpha$, high activity levels are commonly observed in mid M dwarfs, even for only moderate rotating stars; however,
H$\alpha$ activity declines beyond spectral type M7 followed by a strong drop beyond spectral type M9 \citep{giz00,moh03}.
None of these very late objects beyond M7 is seen at high H$\alpha$  activity levels,
i.e., above log~$L_{\rm H\alpha}$/$L_{\rm bol} = -3.9$, which is the mean level in earlier type M dwarfs, 
especially when considering variability and taking the lower values presented in the literature. 
While 1RXS~J115928.5-524717 clearly shows H$\alpha$ emission \citep[see Fig.\,3 in][]{ham04}, no quantitative analysis of its H$\alpha$ line is presented in their work.
Sample studies indeed indicate that the H$\alpha$ activity saturation limit in very late M dwarfs is also
significantly lower than those in earlier type M dwarfs.
However, this decline does not seem to be present for the observed X-ray activity, and the opposite trend is seen in the radio emission from these objects \citep{ber08b}. 
Even the ultracool dwarfs show the same high activity levels 
up to the saturation level as observed for more massive stars (log~$L_{\rm X}$/$L_{\rm bol} \approx -3$).
This could indicate that coronal and chromospheric activity levels do not strictly co-evolve in the regime of ultracool dwarfs.
On the other hand, strong magnetic fields (Bf\,$>$\,1kG) are found in all investigated objects of spectral type M7 and beyond \citep{rei07},
and the strongest magnetic fields (Bf\,$>$\,2kG) are also associated with the most X-ray active stars.
This suggests a correlation - as naturally expected- between magnetic field strength and X-ray activity.

Concerning dynamo mechanisms and their efficiencies and considering that the stellar radii are in a narrow range around 0.1\, R$_{\sun}$ for our sample stars \citep{cha00}, 
the measured $Vsini$ values lead to maximum periods of a few hours for the faster rotators and 
to less than one day even for the slowest rotators. This is significantly shorter than the convective turnover times
expected to be of the order of several ten up to hundred days.
Consequently, the Rossby number (Ro=P/$\tau_{c}$), whose inverse describes dynamo efficiency, is small in all ultracool dwarfs and thus
all stars should be in the saturated (or even super-saturated for the fastest rotators)
regime of magnetic activity, an expectation obviously contradicted by the X-ray observations.
Furthermore, rotation appears to be virtually uncorrelated with H$\alpha$ activity in ultracool dwarfs \citep{reid02}, 
and it does not seem to be the dominant factor determining the X-ray activity level, again in contrast to more massive stars.
Since these correlations refer to an $\alpha\Omega$ dynamo, the X-ray observations support the presence of an alternative dynamo mechanism (e.g. $\alpha^{2}$, turbulent), 
which operates in a fully convective stellar interior, but favour a scenario where activity does not solely depend on rotation.
On the other hand, the saturation level, i.e., the maximum activity level in log~$L_{\rm X}$/$L_{\rm bol}$, appears to be similar in very low-mass and in higher mass stars.
This would then require saturation to be basically independent of the underlying interior structure or dynamo mechanism and to depend 
only on the amount of magnetic flux generated and propagating to the stellar surface where it then leads to the observed activity phenomena.

While the sample size is too small to derive definite conclusions on the overall coronal properties in very low-mass stars, 
the available data clearly confirm that quasi-quiescent coronae exists in many, if not all, late-type stars down to the hydrogen burning mass limit.
The detected stars are moderately to highly active, and their highest X-ray activity levels are comparable to those observed for more massive stars.
Furthermore, the well known trend of higher coronal temperatures towards more active stars is apparently also present in ultracool dwarfs and objects 
with the strongest magnetic field are also the most X-ray active ones. 
Altogether, these findings support the hypothesis that fully convective low-mass stars and solar-like stars exhibit similar overall coronal properties,
regardless of their different interior structures, and that X-ray activity rather depends on the amount of generated magnetic flux
but not on the mechanism responsible for its creation.
Future X-ray observations of ultracool dwarfs are highly desirable to extend and deepen our understanding of magnetic activity 
and coronae in the regime of the coolest stars at the end of the main sequence.

\section{Summary and conclusions}
\label{sum}

   \begin{enumerate}
\item We have clearly detected quasi-quiescent X-ray emission from the ultracool dwarf 1RXS~J115928.5-524717 (spectral type M9) at soft X-ray energies.
The derived X-ray luminosity of about $L_{\rm X} = 1.0 \times 10^{26}$~erg/s in the 0.2\,--1.0~keV band leads
to an activity level of log~$L_{\rm X}$/$L_{\rm bol}\approx -4.1$, pointing to a moderately active star.
It is still relatively poorly studied, however, it is a promising target to be investigated in greater detail  at other wavelengths.

\item The X-ray emitting coronal plasma is best described by at least two temperature components, one relatively cool (2\,MK) and a slightly hotter component (6\,MK). These temperatures are 
also typical for stars of higher mass with a similar activity level. 
We find that the correlation between X-ray activity level and average coronal temperature apparently also holds for ultracool dwarfs.

\item The repeated X-ray detection of very low mass stars suggests that magnetic activity and a stable coronae are quite common phenomena down to the ultracool end of the main sequence. Remarkably, these objects exhibit trends similar to more massive stars in their coronal X-ray properties, despite their different interior structure.
The detected stars are quite active (in $L_{\rm X}$/$L_{\rm bol}$) but overall X-ray faint (in $L_{\rm X}$)
and were mainly detected since they are nearby, thus they are probably only the 'tip of the iceberg' concerning X-ray activity in ultracool dwarfs.
 
   \end{enumerate}

\begin{acknowledgements}
This work is based on observations obtained with XMM-Newton, an ESA science
mission with instruments and contributions directly funded by ESA Member
States and the USA (NASA). J.R. acknowledges support from DLR under 50QR0803.

\end{acknowledgements}

\bibliographystyle{aa}
\bibliography{11224}

\end{document}